\documentstyle[12pt,titlepage]{article}
\input epsf

\font\small=cmr8 scaled \magstep0

\font\medio=cmr10 scaled \magstep2
\outer\def\beginsection#1\par{\medbreak\bigskip
      \message{#1}\leftline{\bf#1}\nobreak\medskip
\vskip-\parskip
      \noindent}

\def\laq{\raise 0.4ex\hbox{$<$}\kern -0.8em\lower 0.62
ex\hbox{$\sim$}}
\def\gaq{\raise 0.4ex\hbox{$>$}\kern -0.7em\lower 0.62
ex\hbox{$\sim$}}

\def\beq{\begin{equation}}
\def\eeq{\end{equation}}
\def\bea{\begin{eqnarray}}
\def\eea{\end{eqnarray}}

\def \pa {\partial}
\def \ra {\rightarrow}

\def \la {\lambda}

\def \b {\beta}
\def \a {\alpha}

\def \da {\delta}
\def \ep {\epsilon}
\def \r {\rho}

\begin{document}
\bibliographystyle {unsrt}

\titlepage
\begin{flushright}
CERN-TH/96-165\\
hep-th/9607126 \\
\end{flushright}
\vspace{20mm}
\begin{center}
{\bf SINGULARITY AND EXIT PROBLEMS} \\
\vskip 0.5 cm
{\bf IN TWO-DIMENSIONAL STRING COSMOLOGY}

\vspace{10mm}

M. Gasperini\footnote{Permanent address: 
{Dip. di Fisica Teorica, Un. di Torino, 
Via P. Giuria 1, 10125 Turin,
Italy.}} and G. Veneziano\\
{\em Theory Division, CERN, CH-1211 Geneva 23, Switzerland} \\
\end{center}
\vspace{10mm}
\centerline{\medio  Abstract}

\noindent
A broad class of two-dimensional loop-corrected dilaton 
gravity models exhibit cosmological solutions that
interpolate  between the string perturbative vacuum and a
background with asymptotically  flat metric and linearly
growing dilaton. The curvature singularities of the
corresponding tree-level solutions are smoothed out, but no 
branch-change occurs.  Thus, even in the presence of a 
non-perturbative potential, the system is not attracted by
physically interesting fixed points with constant dilaton, and 
the exit problem of string cosmology persists. 

\vspace{15mm}

\vfill
\begin{flushleft}
CERN-TH/96-165\\
June 1996 
\end{flushleft}

\newpage

The scale factor duality  of the string effective action
\cite {1}  has recently motivated the study of a class of
cosmological models, in which the Universe starts evolving
from the string perturbative vacuum through an initial,
pre-big bang phase \cite{2,2a} having ``dual" kinematic
properties with respect to those of standard cosmology. 
This initial phase is characterized by an
accelerated growth of the curvature and of the string
coupling, so that the transition to the post-big
bang decelerated evolution is expected to occur in the region
of high curvature and/or strong coupling. Such
a transition cannot be consistently described in the context
of the lowest-order string effective action \cite{3} (unless
one adopts a radical quantum cosmology approach, in which
the transition is described as a scattering process between
asymptotic $|{\rm in} \rangle$ and $|{\rm out}\rangle$ 
states in minisuperspace \cite{4}, thus neglecting details of
the transition region). And in fact, available examples of
non-singular backgrounds, describing a smooth transition
between pre- and post-big bang configurations, make use of
a two-loop non-local dilaton potential \cite{2,5}, or are
formulated as exact conformal field theories \cite{6},  
which automatically take all higher-derivative corrections
into account.

The importance of loop corrections for implementing
non-singular string cosmology models has been recently
emphasized by Easther and Maeda \cite{7}. By extending
previous work on one-loop superstring cosmology \cite{8}, they
have found non-singular four-dimensional solutions that
interpolate smoothly between an intial string phase and a final
era of the Friedmann-Robertson-Walker type, with constant
dilaton and decreasing curvature. Similar results have been
recently obtained also by Rey \cite{9}, working in the context of
the so-called CGHS model of two-dimensional dilaton gravity
\cite{10}, with the one-loop trace anomaly term supplemented
by a local, covariant counterterm in order to preserve 
a useful classical symmetry \cite{11,12}. 
By exploiting such a 
symmetry to generalize the classical solutions, it has been
shown \cite{9} that the curvature and dilaton singularities
of the tree-level pre-big bang background are regularized by
the quantum one-loop corrections. However, in that example 
a limited number of conformal scalar fields ($N<24$) 
must be assumed,  corresponding to a negative contribution to 
the one-loop anomaly term. This is
known to lead to gravitational instabilities, and to the
emission of negative-energy Hawking radiation \cite{12}. Also,
the curvature is bounded but the dilaton keeps growing,
asymptotically, so that higher-loop corrections cannot be
neglected.

In this paper we present a different class of solutions of the CGHS
model, in which the curvature singularity of the tree-level
description is smoothed out by the one-loop terms without
spoiling the physical requirement $N>24$. The dilaton still
evolves monotonically but, unlike in the case discussed in
\cite{9}, the semiclassical back reaction of the produced
gravitational radiation grows in time and may become of the
same order as the one-loop terms, irrespective of the initial
density. However, without some mechanism implementing  a
change of branch of the solution, the growth of the dilaton
can neither be stopped by this back reaction nor by the
effects of a non-perturbative dilaton potential.  

We start considering the one-loop effective action for a
two-dimensional model of dilaton gravity, coupled to $N$
conformal matter fields $f_i$,
\beq
S=\int d^2x\sqrt{-g}\left[-e^{-\phi}\left(R+(\nabla
\phi)^2+\Lambda\right)+{1\over 2}\sum_{i=1}^N (\nabla
f_i)^2+
{k\over 2}\left(R\nabla^{-2}R+\ep\phi R\right)\right] .  
\label{1}
\eeq
Notations: $k=(N-24)/24$, $\nabla$ is the covariant gradient
operator, $g_{00}=+1$, and $R_{\mu\nu\a}\,\,^\b= \pa_\mu
\Gamma_{\nu\a}\,^\b -...$. With our conventions, the dilaton is
related to the effective string coupling $g_s$ by
$e^\phi=g_s^2$. The first contribution proportional to $k$ in eq.
(\ref{1}) corresponds to the usual trace anomaly, the second one
(parametrized by $\ep$) is a local counterterm that one is free
to add to the definition of the model. The case $\ep=0$
reproduces the original CGHS model \cite{10}, the case $\ep=1$ is
the conformal-invariant model considered in \cite{11,12}. 

We are looking for exact cosmological solutions of the above
action with $\Lambda=f_i=0$ (as in \cite{9}), by keeping for the
moment both $k$ and $\ep$  arbitrary (homogeneous
cosmological solutions with non-vanishing $\Lambda$, $f=f(t)$
and $\ep=1$ have already been discussed in \cite{13}). We shall
work in the cosmic time gauge, the most appropriate to
cosmological applications. To this aim we parametrize the
two-dimensional metric in terms of the scale factor $a(t)$ and of
the lapse function ${\cal N}(t)$ as
\beq
ds^2={\cal N}(t) dt^2-a^2(t) dx^2 , ~~~~~a=e^\b , ~~~~~\b=\b(t)
.  \label{2}
\eeq
The action, modulo total derivatives, can thus be rewritten as
\beq
S=\int{dx^2\over {\cal N}}e^\b\left[-\left(e^{-\phi}\right)\dot{}
~(2\dot \b -\dot\phi)+k\dot\b(\ep \dot\phi-2\dot\b)\right] ,
\label{3}
\eeq
where a dot denotes differentiation with respect to cosmic
time $t$. 

We shall vary the action with respect to $\cal N$ and $\b$, 
fixing
the gauge to ${\cal N}=1$. The first variation gives the
Hamiltonian constraint:
\beq
\left(e^{-\phi}\right)\dot{}~(2\dot \b -\dot\phi)=
k(\ep \dot\phi\dot\b -2\dot \b^2) .
\label{4}
\eeq
The second leads to the spatial components of the gravi-dilaton
tensor equations, which can be integrated immediately to give
\beq
{d\over dt}\left(-e^{-\phi}+{k\over 2}\ep\phi -2
k\b\right)= {e^{-\b}\over t_0}  
\label{5}
\eeq
($t_0$ is an integration constant). By eliminating 
$\left(e^{-\phi}\right)\dot{}$ and $\dot \b$ through eq.
(\ref{5}), the constraint (\ref{4}) reduces to
\beq
\dot \phi^2\left[e^{-2\phi} +k(\ep-2)e^{-\phi}+{k^2\ep^2\over
4} \right]={e^{-2\b}\over t_0^2} .
\label{6}
\eeq
The square root of this equation, combined with eq. (\ref{5}),
then leads to the system of coupled first-order equations
\bea
2k\dot \b&=&-{e^{-\b}\over t_0} \pm {e^{-\b}\over
t_0}\left(e^{-\phi}+{k\ep\over 2}\right)
\left[e^{-2\phi} +k(\ep-2)e^{-\phi}+{k^2\ep^2\over
4} \right]^{-1/2} ,
\nonumber \\
\dot\phi&=&\pm {e^{-\b}\over t_0}
\left[e^{-2\phi} +k(\ep-2)e^{-\phi}+{k^2\ep^2\over
4} \right]^{-1/2} ,
\label{7}
\eea
from which, setting $e^\phi=g_s^2$,
\beq
{d\b\over d g_s^2}={1\over 2k g_s^4}\left(1+{k\over 2}\ep
g_s^2 \mp \sqrt{1 +k(\ep-2)g_s^2+{k^2\ep^2\over
4}g_s^4 }~~\right) .
\label{8}
\eeq

For $\ep=1$ we now easily recover the two branches of the
exact solution presented in \cite{9}, characterized, 
respectively, 
by $\b=\ln g_s$ and by $k\b=-g_s^{-1}$, namely $2\b=\phi$ and
$k\b e^{\phi/2}=-1$. Singularities are avoided, in this solution,
only for $k<0$, as can easily be seen from eqs.
(\ref{7}) by noting that, for $\ep=1$ and $k>0$, both $\dot\phi$
and $\dot\b$ diverge at $g_s^2=2/k$.

The case $\ep=1$, however, is only the particular limiting case of
the condition $\ep \geq 1$, under which eq. (\ref{8}) provides
real solutions for $\b (g_s)$. For $\ep >1$, we obtain from
(\ref{8}) the general integral
\bea
\b (g_s^2)&=& \b_0-{1\over 2kg_s^2} \left(1\mp
\sqrt{1 +k(\ep-2)g_s^2+{k^2\ep^2\over
4}g_s^4 }~~\right) + {\ep\over 4}\ln g_s^2 \mp\nonumber\\
&\mp& {|k|\ep \over 4 k}\sinh^{-1}
\left[k^2\ep^2g_s^2/2+k(\ep-2)\over 2|k|\sqrt{\ep-1}\right]
\pm{\ep-2\over 4} \sinh^{-1}
\left[k(\ep-2)g_s^2+2\over 2|k|g_s^2\sqrt{\ep-1}\right] , 
\label{9}
\eea
where $\b_0$ is an integration constant. We shall consider, in
this paper, the physical case $k>0$ (i.e. $N>24$), and we shall
concentrate on the upper branch of the solution, the one that
reduces asymptotically, for $t \ra - \infty$, to the tree-level,
superinflationary pre-big bang solution \cite{2} $a \sim
(-t)^{-1}$, $\phi \sim -2 \ln (-t)$ (in the other branch
$\dot\phi$ diverges as $t\ra -\infty$). 
For positive values $k$ can be
absorbed into $g_s^2$ (with a redefinition of the integration
constant $\b_0$); in addition, for $\ep>1$, the dilaton is a
monotonic function of cosmic time (see eq. (\ref{7})). The general,
upper branch solution with $k>0, \ep>1$ can thus be given as a
function of the monotonic coupling parameter
$g_s^2(t)=\exp[\phi(t)]$ as:
\beq
a(g_s)=e^\b=e^{\b_0}\left|{ g_s^2\over \ep(r+g_s^2) +\ep
-2 }\right|^{\ep/4}\left|2r+2 +(\ep-2)g_s^2\over
g_s^2\right|^{(\ep-2)/4} \exp\left(r-1 \over 2g_s^2\right), 
\label{10}
\eeq
\beq
\dot\phi(g_s) ={g_s^2\over a t_0 r} ~~,  ~~~~~~~~~~~~
2\dot\b ={1\over a t_0 r}\left(1+{\ep\over 2}g_s^2-r\right) ,
\label{11}
\eeq
where
\beq
r(g_s)=\sqrt{1+(\ep-2)g_s^2+{\ep^2\over  4}g_s^4}
\label{12}
\eeq
and $g_s(t)$ is given implicitly by
\beq
{t\over t_0}= \int {dg_s^2\over g_s^4} r(g_s)~a(g_s) . 
\label{13}
\eeq

In this branch, the evolution of the scale factor is monotonic, 
as $d\b/dg_s^2>0$ (see eq. (\ref{8}).  
Asymptotically, at $t \ra -\infty$ and $g_s\ra 0$, we find from
eqs. (\ref{10}) and (\ref{13}) that $a_-(g_s)\sim g_s$ and
$g_s^2=e^\phi\sim (-t)^{-2}$. At  $t \ra +\infty$, $g_s \ra
\infty$, the scale factor approaches a constant, $a_+(g_s)\sim
{\rm const}$, and $g_s^2=e^\phi\sim e^{ct}$, $c={\rm const}$.
The above solution thus describes, for any $\ep>1$, a smooth
transition between the two-dimensional version of the well-known  
\cite{2} dilaton-dominated, pre-big bang inflationary
evolution $a \sim (-t)^{-1}$, $\phi \sim -2\ln(-t)$, and a final
configuration characterized asymptotically by flat space-time
and linearly growing dilaton \cite{14}. The
transition occurs without singularities in $\dot\b$ and
$\dot\phi$ for all $\phi$ ranging from $-\infty$ to $+\infty$, as
can easily be checked from eq. (\ref{7}) and from the fact that
$f(g_s)$ has no real zeros. The scalar curvature
$R=-2(\ddot\b+\dot\b^2)$ is everywhere bounded, approaches
zero at $t\ra \pm \infty$, and reaches a maximum around the
transition region $g_s\sim 1$. The plot of $a(\phi),
~\dot\phi(\phi), ~\dot\b(\phi)$ and $R(\phi)$ is shown in 
Fig. 1 for the particular case $\ep=2$. 

The above class of exact solutions is not completely satisfactory
as an example of regular cosmological backgrounds, however,
because the dilaton keeps growing as $t\ra +\infty$ ($\phi\sim
t$), thus reaching asymptotically a regime in which the
perturbative approximation breaks down (as in the solution
presented in \cite{9}), and the effective action becomes
dominated by higher-loop corrections. Unlike in the
four-dimensional solutions discussed in \cite{7}, there is no
way to obtain from eqs. (\ref{10})--(\ref{13}) a
coupling parameter $e^\phi$ which remains bounded at all times. 

\vskip 1 cm

\centerline{\epsfxsize=4.0in\epsfbox{floop.epsf}}

\vskip 0.5 cm
\noindent
\baselineskip=13 pt
{\small {  
{\bf Fig. 1}. {\em 
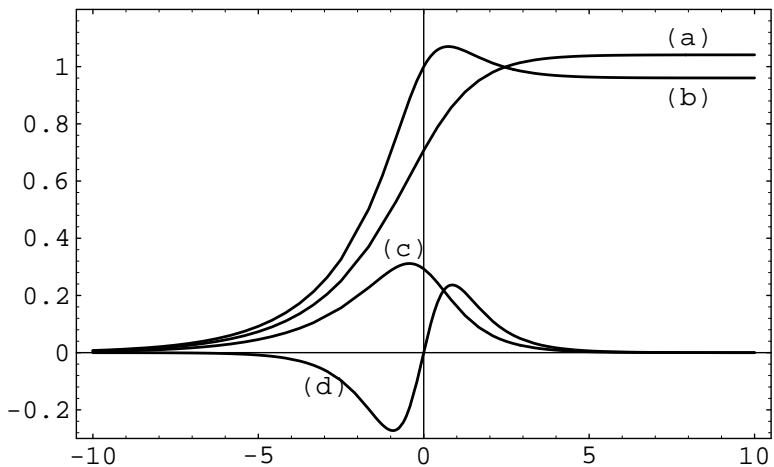
Plot versus $\phi=2\ln g_s$ of (a) the scale factor $a$, (b) the
dilaton growing rate $\dot\phi$, (c) the expansion rate
$\dot\b$, and (d) the scalar curvature
$R=-2(\ddot\b+\dot\b^2)$, for the solution (\ref{10}),
(\ref{11}), in the particular case $\ep=2$. We have chosen
$t_0$ equal to the fundamental string length $\la_s$, and we
have normalized $a$ in such a way that $\dot\phi=1$, in string
units, when $\phi=0$.}}}

\vskip 1 cm

\baselineskip=20pt
In the context of a realistic cosmological model, however, we
should take into account the effect of  a 
non-perturbative dilaton potential $V(\phi)$. Such
a potential, typically required by supersymmetry-breaking
models, is known to go very rapidly to zero at small coupling,
$V(\phi)\sim \exp(-g_s^{-2})$ for $\phi \ra -\infty$, while it
tends to grow with a complicated, in general non-monotonic
behaviour 
 in the opposite, large-coupling limit \cite{15}. Any
potential $V(\phi)$ that grows enough  at large
$\phi$ can thus dominate the
one-loop contributions to the background energy density, 
$ke^\phi\dot\b\dot\phi$, which stay constant at large $\phi$,
and might suppress the asymptotic growth of the dilaton
(different examples of regular two-dimensional backgrounds,
without loop terms but with an appropriate potential, have
been discussed in \cite{16} in the context of models
implementing the limiting curvature hypothesis \cite{17}).

In order to discuss this possibility we add to the action
(\ref{3}) a potential $V(\phi)$,
\beq
S=\int{dx^2\over {\cal
N}}e^\b\left\{e^{-\phi}\left[\dot \phi (2\dot \b
-\dot\phi) -{\cal N}^2V(\phi)\right] 
+k\dot\b(\ep \dot\phi-2\dot\b)\right\} . 
\label{14}
\eeq
The variation with respect to $\cal N, \b$ and $\phi$ provides
the equations (in the gauge ${\cal N}=1$)
\beq
\dot\phi^2-2\dot\b \dot\phi-V=ke^\phi\dot\b(\ep
\dot\phi-2\dot\b),
\label{15}
\eeq
\beq
\left(\ddot\phi +\dot\b \dot\phi\right)
\left(1+k{\ep\over2}e^\phi\right)-2ke^\phi\left(\ddot\b+
\dot\b^2\right)-\dot\phi^2+V=0,
\label{16}
\eeq
\beq
\ddot\phi=\left(\ddot\b+
\dot\b^2\right)\left(1+k{\ep\over2}e^\phi\right)
+{1\over 2}\dot\phi^2-\dot\b\dot\phi+{1\over 2}(V'-V),
\label{17}
\eeq
where $V'=\pa V/\pa \phi$. These equations 
can be exactly solved by the particular (de Sitter-like)
configuration with constant dilaton, $\phi=\phi_0={\rm const}$,
and constant curvature, $\dot \b=H_0={\rm const}$, provided
the potential satisfies, at $\phi=\phi_0$,
\beq
V_0=2kH_0^2e^{\phi_0} ~, ~~~~~~~~
V'_0=-V_0\left({e^{-\phi_0}\over k}+{\ep-2\over 2}\right) 
\label{18}
\eeq
(the second condition is required to satisfy the dilaton
equation (\ref{17})). It is interesting to note that $\phi_0$ is
not necessarily an extremum of $V(\phi)$, if $V_0\not= 0$. The
background energy density, in this context, may thus become
vacuum-dominated even if the scalar field is not at the
minimum of the potential, and this may have interesting
implications for the solution of the cosmological constant
problem. 

Unfortunately, however, such a frozen configuration
$\{\phi_0,H_0\}$ is not a stable fixed point towards which the
background can be attracted, if we consider the branch of the
solution that includes the phase of pre-big bang evolution
from the perturbative vacuum. In fact, by taking the square
root of eq. (\ref{15}), and eliminating $\ddot \phi$ in eq.
(\ref{16}) through eq. (\ref{17}), we find that in the presence of
a potential the two branches are defined by the equations:
\bea
\dot\phi&=&\dot\b\left(1+{\ep\over 2}g_s^2\right)\pm
\sqrt{\dot\b^2r^2+V} ,\label{19}\\
\ddot\b&=&-\dot\b^2+{1\over 2
r^2}\left[\left(\dot\phi^2-V\right)\left(1-{\ep\over2}g_s^2
\right)-V'\left(1+{\ep\over 2}g_s^2\right)\right]
\label{20}
\eea
(again, we have absorbed $k$ into $g_s^2$). The pre-big bang
branch, which reduces to the solution (\ref{10})--(\ref{13}) for
small enough $g_s$ (when $V(\phi)$ becomes negligible),
corresponds to the upper sign in eq. (\ref{19}). Both branches
are satisfied by the constant dilaton and constant curvature
solution (\ref{18}), with $H_0<0$ for the upper branch, and
$H_0>0$ for the lower one. By perturbing eqs. (\ref{19}),
(\ref{20}) around such a configuration, to first order in 
$\da \phi$ and $\da\dot\b$, we 
can easily compute the $2\times 2$ matrix $\cal M$
characterizing the small oscillations of the background, such
that
\beq
\pmatrix{\da \dot\phi \cr \da \ddot\b }= {\cal M}
\pmatrix{\da \phi \cr \da \dot\b }, ~~~~~~
\dot\b= H_0 , ~~~~ \dot\phi_0=0=\ddot \b_0 .
\label {21}
\eeq
A necessary condition for the stability of the point
$\{\phi_0, H_0\}$ is the presence of a negative real part in
both the eigenvalues of $\cal M$, namely ${\rm Tr}~{\cal M}
<0$.  We find that, for the two branches,
\beq
{\rm Tr}~{ \cal M}= \pm |H_0|,
\label{22}
\eeq
so that $\phi=\phi_0$, $\dot \b=H_0$ cannot be an attractor for
a gravi-dilaton background, emerging from the small coupling
regime and  described by the upper branch of the solution for
which ${\rm Tr} ~{\cal M}>0$. This conclusion can be easily
checked by a numeric integration of eqs. (\ref{19}), (\ref{20}),
for any shape of the dilaton potential. 

As another possibility of stopping the dilaton growth, even
without a dilaton potential, we recall that  
the transition between two phases
with different asymptotic vacua, in general, leads to the
parametric amplification of the initial vacuum fluctuations, and
to the production of radiation \cite{18}. In our
two-dimensional class of backgrounds, the contribution of the
radiation energy density ($\rho_r$) to the space-time
curvature, $e^\phi\r_r$, tends to grow exponentially in time
since $e^\phi\r_r\sim e^\phi a^{-2}\ra e^\phi$ for
$t\ra+\infty$. Consequently, for $\ep>1$, the radiation
contribution will eventually become of the same order as that
of the one-loop terms, whose contribution
to the curvature,
$ke^\phi\dot\b\dot\phi$, tends to a constant 
for $t\ra+\infty$. This is to be contrasted with 
the $\ep=1$ case \cite{9}, where $e^\phi \r_r$ goes to a
constant, asymptotically, and remains negligible.

If we take into account the back reaction of this radiation
produced semiclassically, we must add the term $e^\phi \r_r$
to the right hand-side of eq. (\ref{15}), while the remaining
equations (\ref{16}) and (\ref{17}) are left unchanged because
of the particular form of the radiation equation of state,
$p=\r$, in two dimensions. In the absence of a potential the
two branches are then simply defined by the equations
\bea
\dot\phi&=&\dot\b\left(1+{\ep\over 2}g_s^2\right)\pm
\sqrt{\dot\b^2r^2+\r_0e^{\phi-2\b}} ,
\label{23}\\
\ddot\b&=&-\dot\b^2+{\dot\phi^2\over 2r^2}\left(1-{\ep\over 2} 
g_s^2\right)  , 
\label{24}
\eea
where $\r_0$ is a free parameter. Thanks to the one-loop
terms that cancel the radiation contribution, these equations
admit the particular vacuum solution with  constant dilaton
and globally flat space-time,
\beq
\dot\phi=0, ~~~~~ \ddot\b+\dot\b^2=0, ~~~~~
\r_r \equiv \r_0 e^{-2\b}=2 k \dot\b^2, 
\label{25}
\eeq
describing a linear contraction in the upper (pre-big bang) 
branch, 
$\dot\b<0, a\sim (-t), t<0$, 
and linear expansion in the other branch, 
$\dot\b>0, a\sim t, t>0$. However, as in the previous case
with the dilaton potential, this solution is unstable in the
upper branch (as $\da \ddot\b =\pm 2|\dot\b|\da \dot\b$), and
the  particular solution (\ref{25}) cannot be approached by a
background that evolves from the pre-big bang configuration
without branch changing. 

In conclusion, combining these with other \cite{2},  
\cite{5}--\cite{9} examples of smooth transitions for a typical string
cosmology scenario, the following picture seems to emerge. In
the solutions of the low-energy string effective action the
growth of the curvature is unbounded, and prevents a
continuous evolution from the initial accelerated phase to the
present decelerated regime. At small coupling, higher-order
$\a'$ corrections, typically due to finite-size effects and
weighted by the inverse of the string tension, can ``flatten"
the growth of the curvature, leading eventually 
to a constant-curvature, 
de Sitter-like evolution. At higher coupling, quantum
loop corrections seems to be able to induce a ``bounce" of the
curvature, implementing the transition to the decelerated,
decreasing curvature regime. This transition is  
accompanied, in general, by radiation production, whose
semiclassical back reaction may be expected to become
important, and eventually dominate (``reheating").  
The possible residual growth of the
string coupling, 
however, can be permanently stopped by the radiation back
reaction, or by the effects of non-perturbative
self-interactions,  only in the phase described by the
decelerated branch with ${\rm Tr} ~{\cal M}<0$. A true ``graceful
exit" from the string  to the standard cosmological phase
thus seems  to require a change of branch of the solution,
possibly implemented by the contribution of higher derivative
terms, which are absent in the example discussed here.

Much work is still needed, of course, to clarify all the details of
this cosmological scenario. A full four-dimensional analysis, in
particular, should complement this two-dimensional
discussion. These preliminary results provide
useful hints for future investigations and  for implementing, in
a string theory context, a consistent description of the Universe
which starts evolving from the flat and cold  perturbative
vacuum, and ends up in the present matter-dominated state,
with all the relic aspects of the big bang explosion. 

\vskip 2 cm 

\section*{Acknowledgements}

We thank Adel Bilal, Curtis Callan, Soo-Jong Rey and Jorge Russo for
useful discussions on two-dimensional dilaton gravity. 
This work was supported in part by the ``Human
Capital and Mobility  Program" of 
the European Commission, under the 
contract No. ERBCHRX-CT94-0488.

 \end{document}